\documentclass[iop, apj]{emulateapj}

\usepackage[cp1251]{inputenc}
\usepackage[english]{babel}
\RequirePackage[T2A]{fontenc}
\usepackage{amsmath}
\usepackage{amsfonts}
\usepackage{amssymb}
\usepackage{graphicx}

\shorttitle{Breakout reconnection observed by the TESIS EUV telescope}
\shortauthors{A.~Reva, A.~Ulyanov, S.~Shestov, S.~Kuzin}

\begin{document}
\title{Breakout reconnection observed by the TESIS EUV telescope}
\author{Reva A.A., Ulyanov A.S., Shestov S.V., Kuzin S.V.}
\affil{Lebedev Physical Institute, Russian Academy of Sciences}
\email{reva.antoine@gmail.com}

\begin{abstract}
We present experimental evidence of the coronal mass ejection (CME) breakout reconnection, observed by the TESIS EUV telescope. The telescope could observe solar corona up to 2~$R_\odot$ from the Sun center in the Fe~171~\AA\ line. Starting from 2009 April~8 TESIS, observed an active region (AR) that had a quadrupolar structure with an X-point 0.5~$R_\odot$ above photosphere. A reconstructed from the MDI data magnetic field also has a multipolar structure with an X-point above the AR.  At 21:45~UT on April~9, the loops near the X-point started to move away from each other with a velocity of $\approx$~7~km~s$^{-1}$. At 01:15~UT on April~10, a bright stripe appeared between the loops, and the flux in the GOES 0.5--4~\AA\ channel increased. We interpret the loops' sideways motion and the bright stripe as evidence of the breakout reconnection. At 01:45~UT, the loops below the X-point started to slowly move up. At 15:10~UT, the CME started to accelerate impulsively, while at the same time a flare arcade formed below the CME. After 15:50~UT, the CME moved with constant velocity. The CME evolution precisely followed the breakout model scenario.  
\end{abstract}


\section{Introduction}

Coronal mass ejections (CMEs) are giant eruptions of coronal plasma into interplanetary space. CMEs are the result of an energy release in the corona and are one of the major factors that affect space weather \citep{Schwenn2006, Pulkkinen2007}. Investigation of CMEs are important for solar physics and questions of solar-terrestrial connections.

According to the standard CME model, before eruption, the CME has a bipolar magnetic field structure with a prominence above it \citep{Carmichael1964, Sturrock1966, Hirayama1974, Kopp1976}. The prominence is surrounded by circular magnetic field lines. Under the prominence, there is a current sheet in which reconnection happens. Plasma outflow from the reconnection region pushes the prominence up and the CME erupts.  

In a bipolar configuration, when a CME erupts, all overlying loops should open up (that is, reconnect to infinity) and free the way for the CME. \citet{Aly1984} and \citet{Sturrock1991} showed that, in this case, the final open magnetic configuration has more energy than the initial closed configuration. So, with respect to energy, eruption is impossible. This contradiction is known as the Aly-Sturrock limit.

\begin{figure*}[!t]
\centering
\includegraphics[width = 0.8\textwidth]{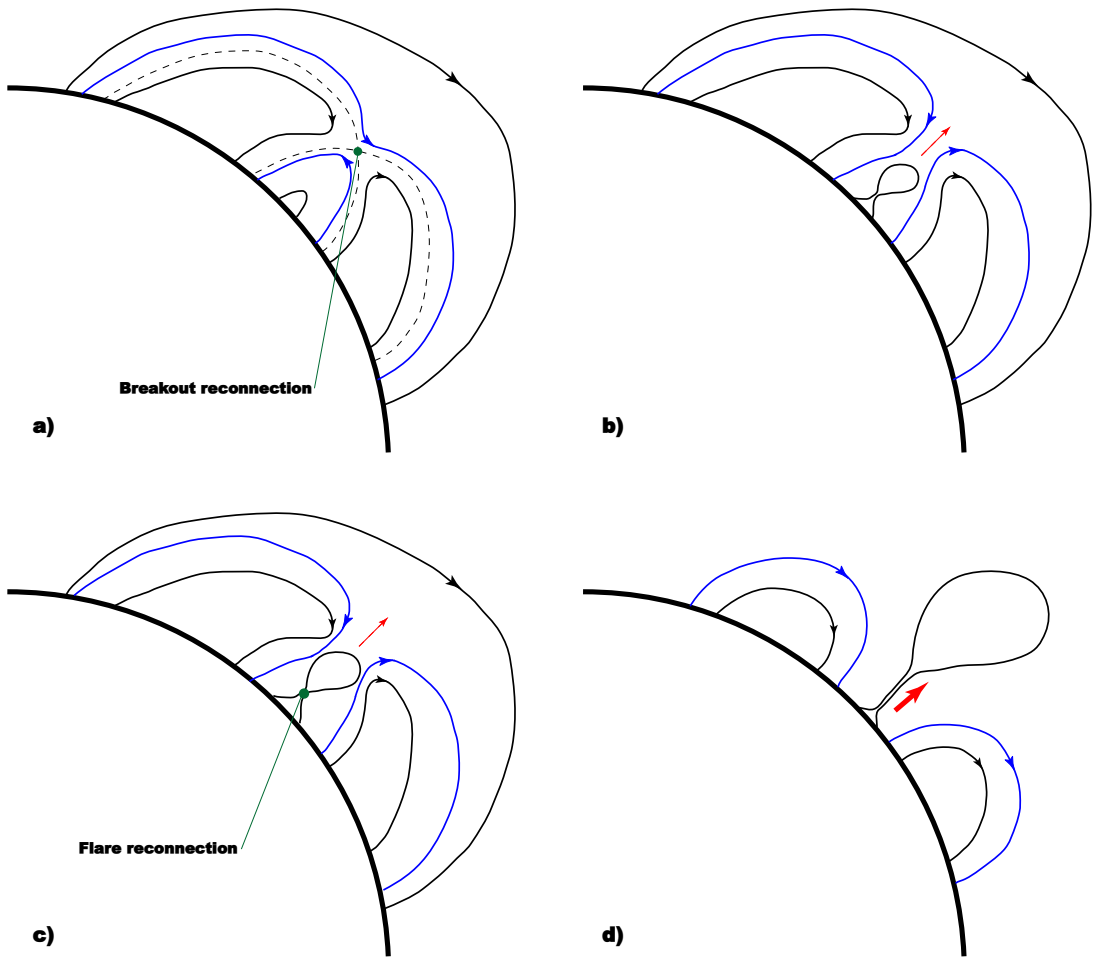}
\caption{Breakout model: a) breakout reconnection; b) CME slowly moves up; c) flare reconnection; d) impulsive acceleration. Thick black line---solar surface; thin black lines---magnetic field lines; blue lines---reconnecting magnetic field lines. }
\label{F:Breakout}
\end{figure*}

To resolve the Aly-Sturrock limit, \citet{Antiochos1999} proposed a `breakout model' of CME, in which overlying loops remain closed after the eruption. In the breakout model,  an arcade is confined by a quadrupolar magnetic structure (see Figure~\ref{F:Breakout}a). The arcade experiences slow shearing motion. The shearing motion pumps into the arcade non-potential magnetic energy, which is needed for an eruption. The reconnection in the X-point of the quadrupolar structure is so slow that it allows to store magnetic energy in the sheared arcade. Eventually, the X-point stretches into the current sheet, and overlying loops reconnect (see Figure~\ref{F:Breakout}a). This external reconnection---which is called \emph{the breakout reconnection}---removes overlying flux above the X-point and sends it to the side. The breakout reconnection relieves magnetic tension and allows the sheared structure to rise.  The CME slowly moves up (see Figure~\ref{F:Breakout}b). The motion stretches the current sheet below the CME, and \emph{flare reconnection} occurs (see Figure~\ref{F:Breakout}c). Plasma outflow from the reconnection region pushes the CME up, and the CME impulsively accelerates and erupts (see Figure~\ref{F:Breakout}d). In the breakout model, the overlying loops remain closed after eruption and, therefore, the Aly-Sturrock limit is not violated.

The breakout model has been extensively studied with numerical magnetohydrodynamic (MHD) simulations. The first complete simulation of breakout CME evolution was performed by \citet{MacNeice2004}. The first 3D simulation of breakout CME was performed by \citet{Lynch2005}. \citet{Karpen2012} performed a high-resolution MHD simulation of the breakout model and reproduced all aspects of the model described in the previous paragraph.

Several events showed indirect evidence of the breakout model: quadrupolar magnetic field structure of the Bastille day flare \citep{Aulanier2000}; `EIT crinckles' \citep{Sterling2001}; dimmings before the CME \citep{Sterling2004Feb}; CME quadrupolar magnetic structure inferred from SXT images \citep{Sterling2004Oct}; timing of brightenings during CME \citep{Gary2004}; and prominence location relative to hotter plasma in the CME \citep{Reva2014}. \citet{Manoharan2003, Aurass2011, Aurass2013} reported radio sources above null points of the CME quadrupolar structure before eruption, which the authors interpreted as evidence of breakout reconnection. However, despite the success of the breakout model, its key element---the breakout reconnection---was never directly observed.

The breakout magnetic structure is expected to be large. Its X-point should be located at a distance of 1.2--2~$R_\odot$ from the Sun's center (we will call this distance range `far corona'). To confirm the breakout model, we need observations in the far corona. Although instruments that could observe far corona existed---LASCO/C1 coronograph \citep{Zhang2001}, Siberian solar radio telescope \citep{Alissandrakis2013}, Mauna Loa Observatory \citep{Bemporad2007}, TESIS EUV telescopes \citep{Reva2014}, and SWAP EUV telescope \citep{Seaton2013}---they were never used to test the breakout model. 

One of the TESIS EUV telescopes observed far corona in 171~\AA\ \citep{Kuzin2011}. Solar corona images in the 171~\AA\ line show individual coronal loops, and therefore, could reveal coronal magnetic structure. In this work, we present TESIS CME observations in the 171~\AA\ line, which, for the first time, directly showed evidence of the breakout reconnection.

\section{Experimental Data}

We analyzed CME that occurred on 2009 April 10. We used the TESIS 171~\AA\ telescope data to study CME behavior below 2~$R_\odot$, and the LASCO/C2,~C3 data above 2~$R_\odot$. We also used \textit{STEREO-A} EUVI images to observe CME evolution behind the limb, and COR2 images to study the CME kinematics. We used MDI data for the reconstruction of the magnetic field.

TESIS is an instrument assembly that observed the solar corona in EUV  and soft X-ray. It worked on board the \textit{CORONAS-PHOTON} satellite in 2009 \citep{Kotov2011}. TESIS included an EUV telescope, which built solar corona images in the Fe~171~\AA\ line with 1.7$^{\prime\prime}$ angular resolution and 1$^{\circ}$ field of view. For the Fe~171~\AA\ telescope, a special observational program was developed, in which the telescope built images of far corona \citep[for more details, see][]{Reva2014}. From April~8 to 10, TESIS worked in the far corona mode with a varying cadence of 10, 20, and 30 minutes. We used TESIS  171~\AA\ telescope data to study CME behavior below 2~$R_\odot$.

LASCO is a set of coronographs that observe solar corona in white light \citep{Brueckner1995}. LASCO works on the \textit{SOHO}  satellite \citep{Domingo1995}. LASCO coronographs cover different distance ranges: C1 1.1--3~$R_{\odot}$; C2 2--6~$R_{\odot}$; and C3 4--30~$R_{\odot}$. LASCO/C1 stopped operating in 1998, and today corona below 2~$R_\odot$ is a `blind zone' for LASCO. We use LASCO data to complement TESIS data at distances above 2~$R_\odot$.

\textit{STEREO} is a set of two satellites that move along the Earth's orbit: \textit{STEREO-A} moves ahead of the Earth and \textit{STEREO-B} moves behind it \citep{Howard2008}. \textit{STEREO} carries Extreme UltraViolet Imager (EUVI), which builds solar corona images in 171, 195, 284, and 304~\AA\ lines. We use \textit{STEREO-A} EUVI images to observe early CME evolution behind the solar limb.

\textit{STEREO} also carries two coronagraphs: COR1 and COR2. COR1 builds corona white light images at a distance range of 1.3--4~$R_\odot$ and COR2 at 2--15~$R_\odot$. Due to the gap in the observations, COR1 did not observed the studied CME and COR2 observed it above 4~$R_\odot$. We use COR2 data to study the CME kinematics.

\begin{figure}[!t]
\centering
\includegraphics[width = 0.45\textwidth]{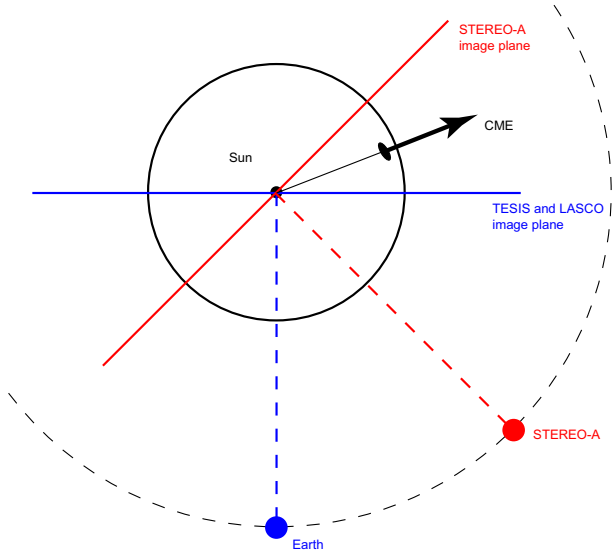}
\caption{\textit{STEREO-A}, Earth, and CME positions.}
\label{F:stereo_position}
\end{figure}

On 2009 April 10 \textit{STEREO-A} was $46^\circ$ ahead of the Earth (see Figure~\ref{F:stereo_position}). The active region(AR) was behind the limb, and the CME trajectory was inclined by approximately the same angle to the TESIS and \textit{STEREO-A} image planes.

The Michelson Doppler Imager (MDI, \citet{Scherrer1995}) works on the \textit{SOHO}  satellite. In a synoptic mode, MDI maps the line-of-sight component of the photospheric magnetic field with a  $4^{\prime\prime}$ resolution and 90-minutes cadence. We used MDI data to reconstruct the magnetic field.

\section{Results}

\subsection{TESIS Observations}

We studied an AR, which appeared above the eastern limb in the northern hemisphere on 2009 March~26, and crossed the western limb on 2009 April~8. The AR could be traced further with \textit{STEREO-A} EUVI telescopes. In 2009, the Sun was in the deep minimum of the solar activity and the analyzed AR was very weak. It did not have any sunspots and was not classified by NOAA.

The TESIS EUV telescope observed this AR in the far corona mode beginning on April 8, 12:00~UT. At this time, the AR was already at the western limb and TESIS could see its far corona structure.

\begin{figure*}[!th]
\centering
\includegraphics[width = 0.95\textwidth]{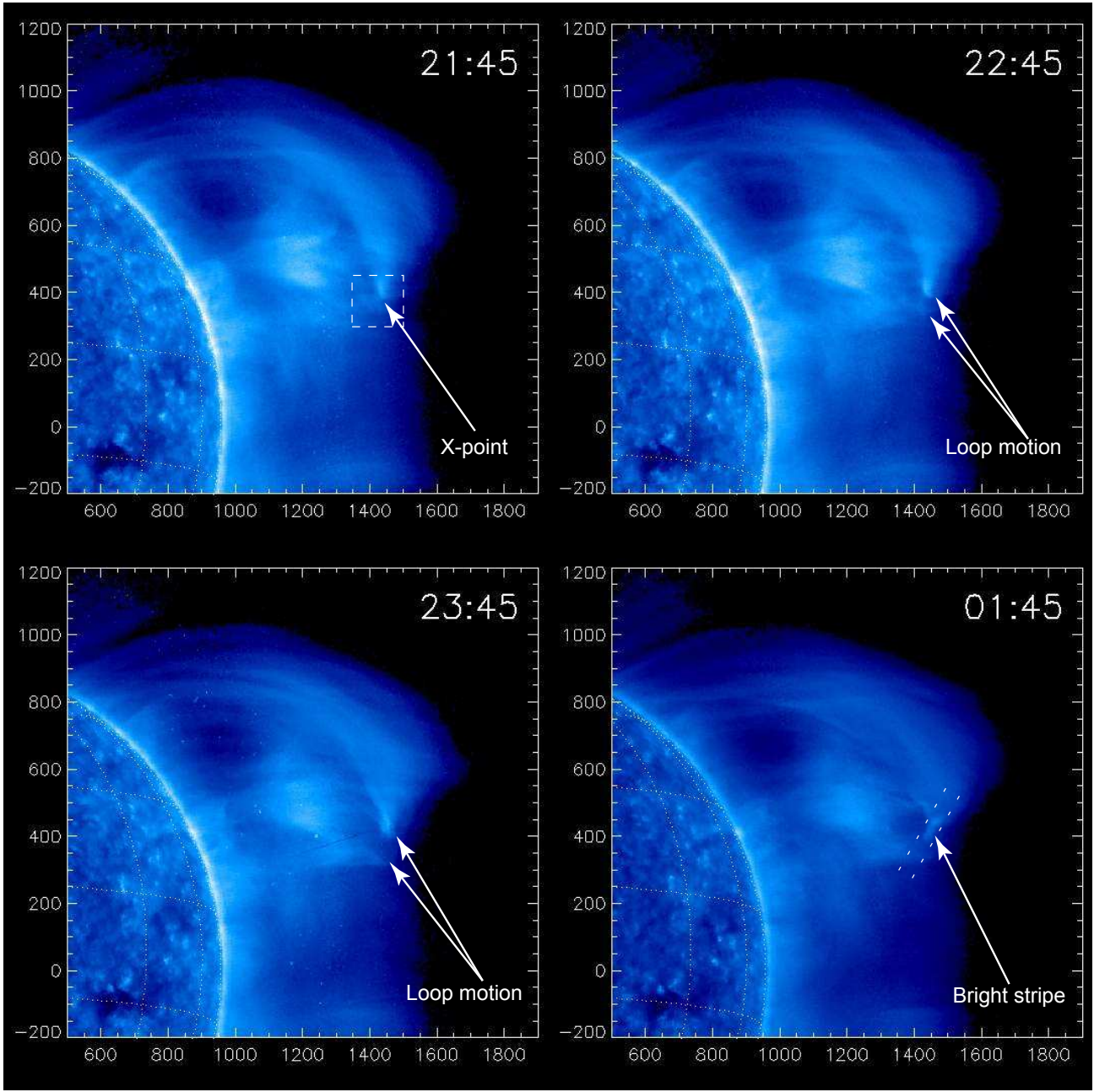}
\caption{Reconnection of the overlying loop. The dashed lines indicate the position and the width of the slices, which were used in the construction of the Figure~\ref{F:tesis_slices}. The dashed rectangle shows the field of view of Figure~\ref{F:X_point}. Images were taken on April 9, 2009. Coordinates are measured in arc seconds.}
\label{F:reconnection}
\end{figure*}

\begin{figure*}[!th]
\centering
\includegraphics[width = 0.95\textwidth]{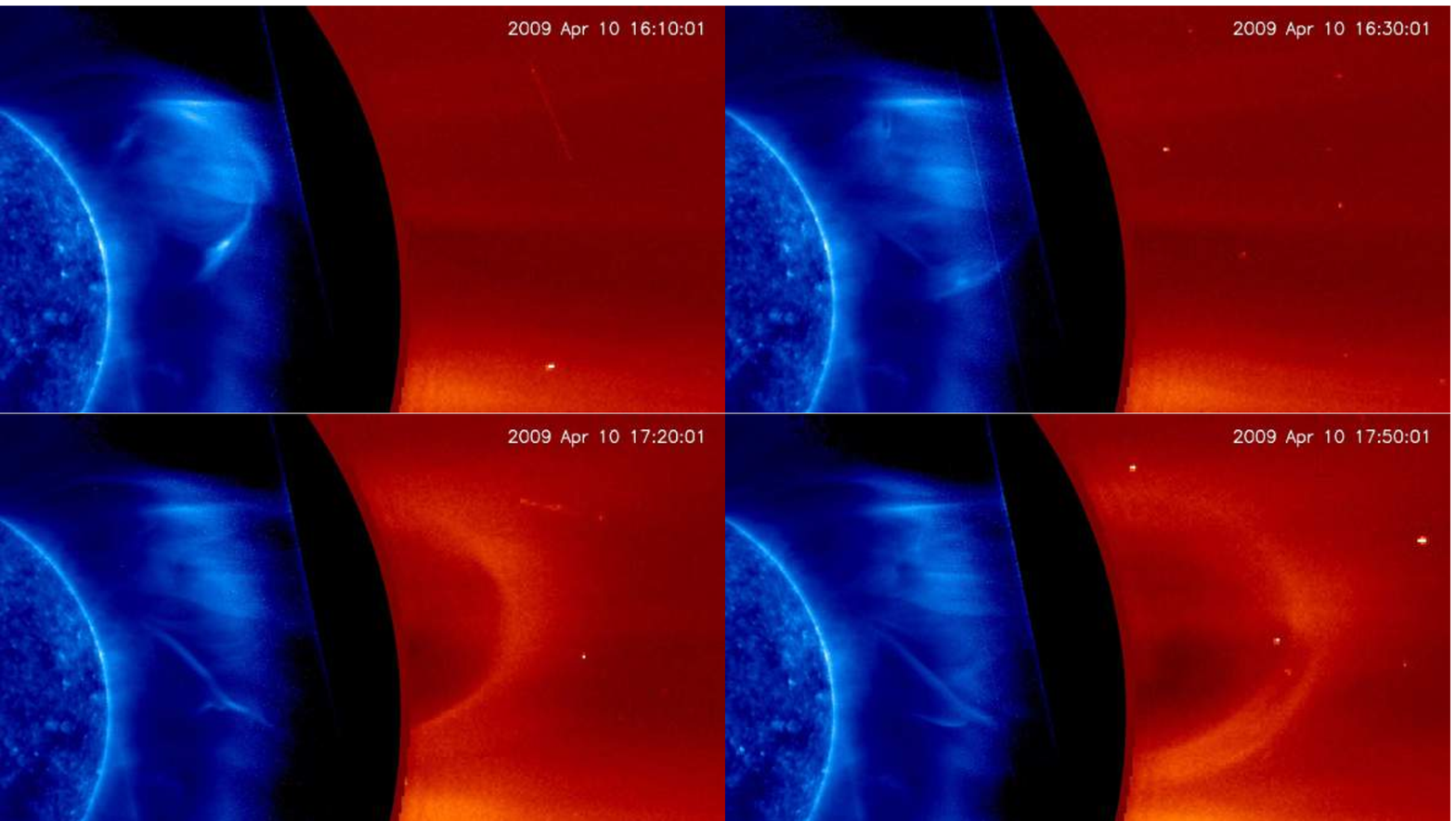}
\caption{CME. Blue: TESIS 171~\AA\ images; red: LASCO/C2 images. The video is available for this figure.}
\label{F:Fe_C2}
\end{figure*}

In TESIS 171~\AA\ images before eruption, the analyzed AR had a quadrupolar structure with a size of $\approx$~750~Mm (see Figure~\ref{F:reconnection}). Although it is hard to distinguish the AR structure on a single image, an inspection of the video for Figure~\ref{F:Fe_C2} allows us to distinguish individual loops, their dynamics, and point them on individual images. The X-point of the magnetic structure was located at the height of $\approx$~0.5~$R_\odot$ ($\approx$~350~Mm) above the photosphere. 

\begin{figure*}[!th]
\centering
\includegraphics[width = 0.95\textwidth]{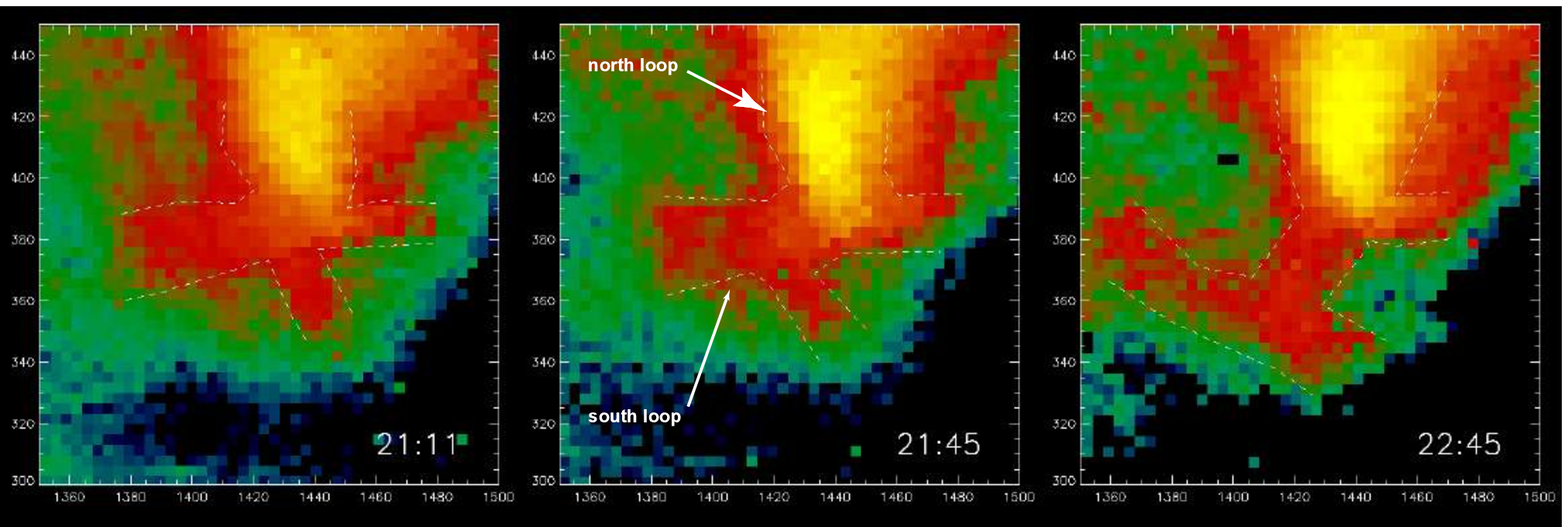}
\caption{Breakout X-point, observed by TESIS (a zoomed version of the Figure~\ref{F:reconnection}). Blue and green correspond to low intensities; red and yellow correspond to high intensities. Contours denote the X-point. Arrows indicate the north and south loops of the quadrupolar structure. Images were taken on 2009 April 9. Coordinates are measured in arc seconds.}
\label{F:X_point}
\end{figure*}

Figure~\ref{F:X_point} shows a zoomed image of the X-point. To highlight the X-point, we used a multicolor color table. Although the inferred X-point is not fully revealed in this Figure, we can see the connected apexes  of the north and the south loops of the quadrupolar system. We can also see that these apexes started to diverge from each other.

On 2009 April 9, at 21:45 UT, loops near the X-point started to move away from each other (see Figure~\ref{F:reconnection} and \ref{F:X_point}).  On 2009 April 10, at 01:15~UT, a bright stripe (80~Mm in length) appeared between the loops (see Figure~\ref{F:reconnection} and the video for Figure~\ref{F:Fe_C2}). We interpreted the loops' motions and the bright stripe as evidence of the breakout reconnection.

\begin{figure}[!b]
\centering
\includegraphics[width = 0.45\textwidth]{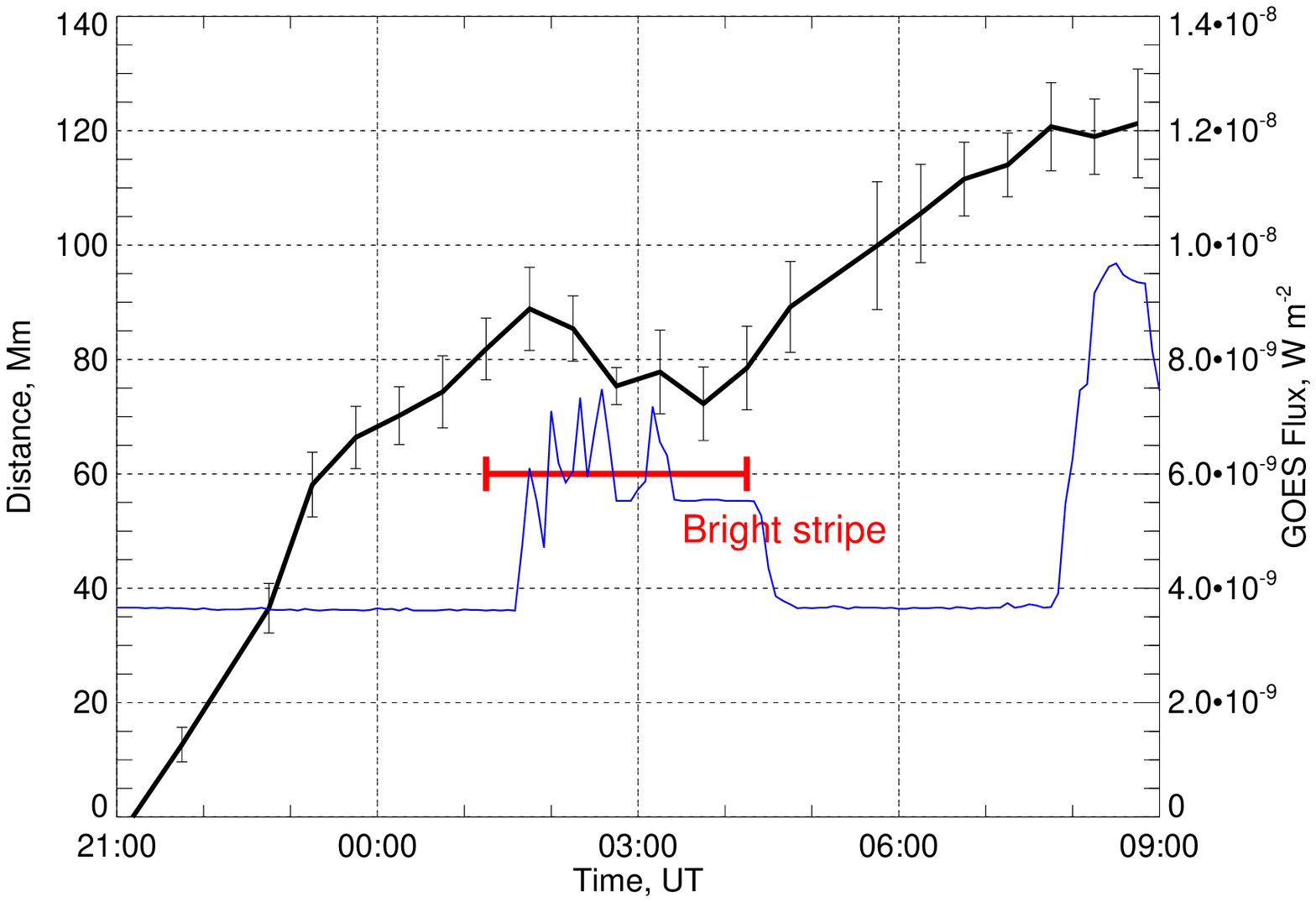}
\caption{Distance between the loops, which formed the breakout X-point. Red indicates the time when the bright stripe appeared and disappeared. Blue line indicates GOES 0.5--4~\AA\ flux.}
\label{F:Loop movement}
\end{figure}

We measured the time dependence of distance between loops, which formed the X-point (see~Figure~\ref{F:Loop movement}). The measurement was carried out manually with a simple point-and-click method. To estimate error bars, we repeated the procedure 25~times. The loops moved away from each other at a speed of $\approx$~7~km~s$^{-1}$. When the bright stripe appeared, the movement stopped for $\approx$~3~hours, and the flux in GOES the 0.5--4~\AA\ channel increased (see~Figure~\ref{F:Loop movement}). When the bright stripe disappeared, the movement continued.

\begin{figure*}[!th]
\centering
\includegraphics[width = 0.95\textwidth]{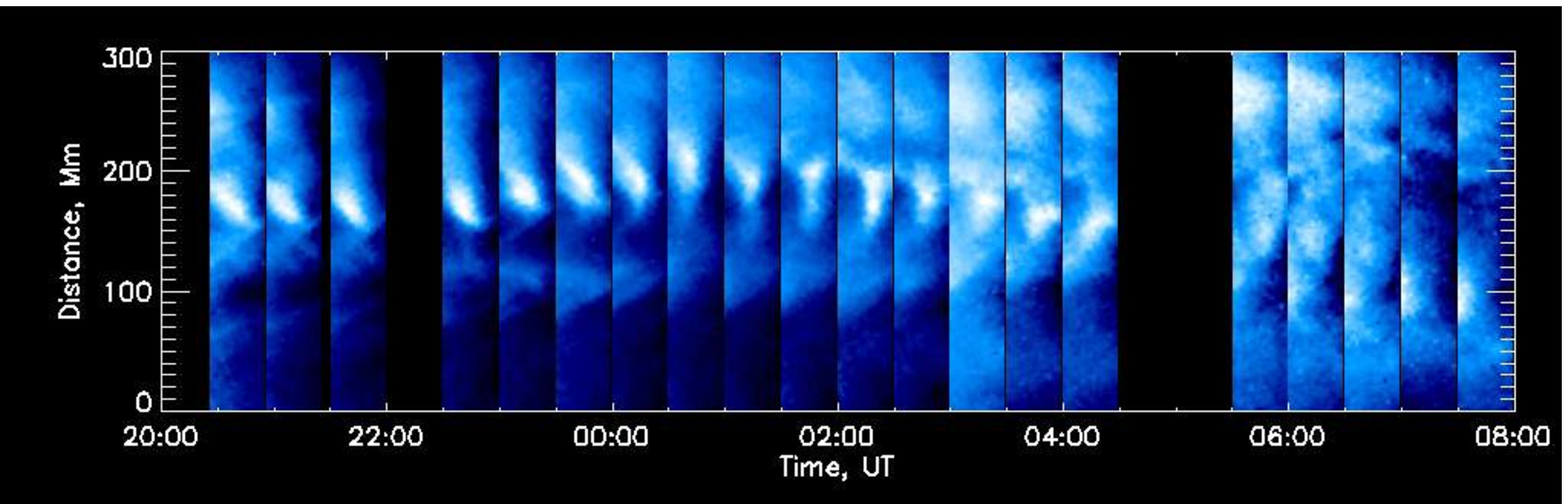}
\caption{Stack plot of the slices, which were cut from TESIS images in the vicinity of the X-point. Position of the slices is indicated on the Figure~\ref{F:reconnection}. The width of the slices is 60$^{\prime\prime}$ (44~Mm).}
\label{F:tesis_slices}
\end{figure*}

To illustrate the loops' motion more clearly, we made a stack plot of slices, cut from the TESIS images in the vicinity of the X-point (see Figure~\ref{F:tesis_slices}). The position of the slices is shown in Figure~\ref{F:reconnection}. During the evolution, the bright stripe slightly moved up. We moved the position of the slices to follow the vertical motion of the bright stripe. 

In Figure~\ref{F:tesis_slices}, the upper bright feature ($y$=190~Mm, $t$=00:00~UT) corresponds to the apex of the north loop, and the bottom bright feature ($y$=110 Mm, $t$=00:00~UT) corresponds to the apex of the south loop. Before 21:45~UT these loops touched each other and formed an X-point. From 21:45 to 01:15~UT, the loops diverged from each other. Both loops started to move simultaneously. We interpret this motion as a restructuring of the magnetic field (evidence of the breakout reconnection). From 01:15 to 04:15~UT, the bright stripe lightens up between the loops. The stripe resembled the current sheet, and we interpret it as another evidence of the breakout reconnection. An apparent converging motion of the bright features (starting from 03:15~UT) is not a loop motion, but a shift of the bright stripe intensity maximum from the north to the south loop.

Starting at 03:15~UT, the structures are more blurred, and it is harder to identify loop apexes. The pause in the loop divergence motion reported above  could be a result of a subjective mistake in the identification of the blurred loop apexes.

The bright stripe intensity on the TESIS images was 160~DNs, the coronal hole intensity was 1010~DNs, and the quiet Sun intensity was 4120~DNs. So, the bright stripe intensity amounted to 15~\% of the coronal hole intensity or 4~\% of the quiet Sun intensity. On the \textit{STEREO-A} images, quiet Sun intensity was 700~DNs but the bright stripe was not seen. Its intensity should be around 30~DNs. Most likely, we did not observe the bright stripe on the EUVI-A images due to its low intensity, or because the corresponding EUVI-A line of sight is on-disk and thus intercepts denser layers.

On 2009 April 10, at 15:00 UT, the CME erupted (see Figure~\ref{F:Fe_C2}). It had a loop-like shape. We measured coordinates of the CME's leading edge on the TESIS and LASCO images. The CME moved  with almost constant velocity of $\approx$~165~km~s$^{-1}$. When the CME erupted, its footpoints were  behind the limb, so we did not see CME acceleration on the TESIS images.

\subsection{Magnetic Field}

We assumed above that the observed structure had a 2D magnetic field configuration, which could be incorrect. In that case, more complex 3D models should be adopted. 

\begin{figure}[!b]
\centering
\includegraphics[width = 0.45\textwidth]{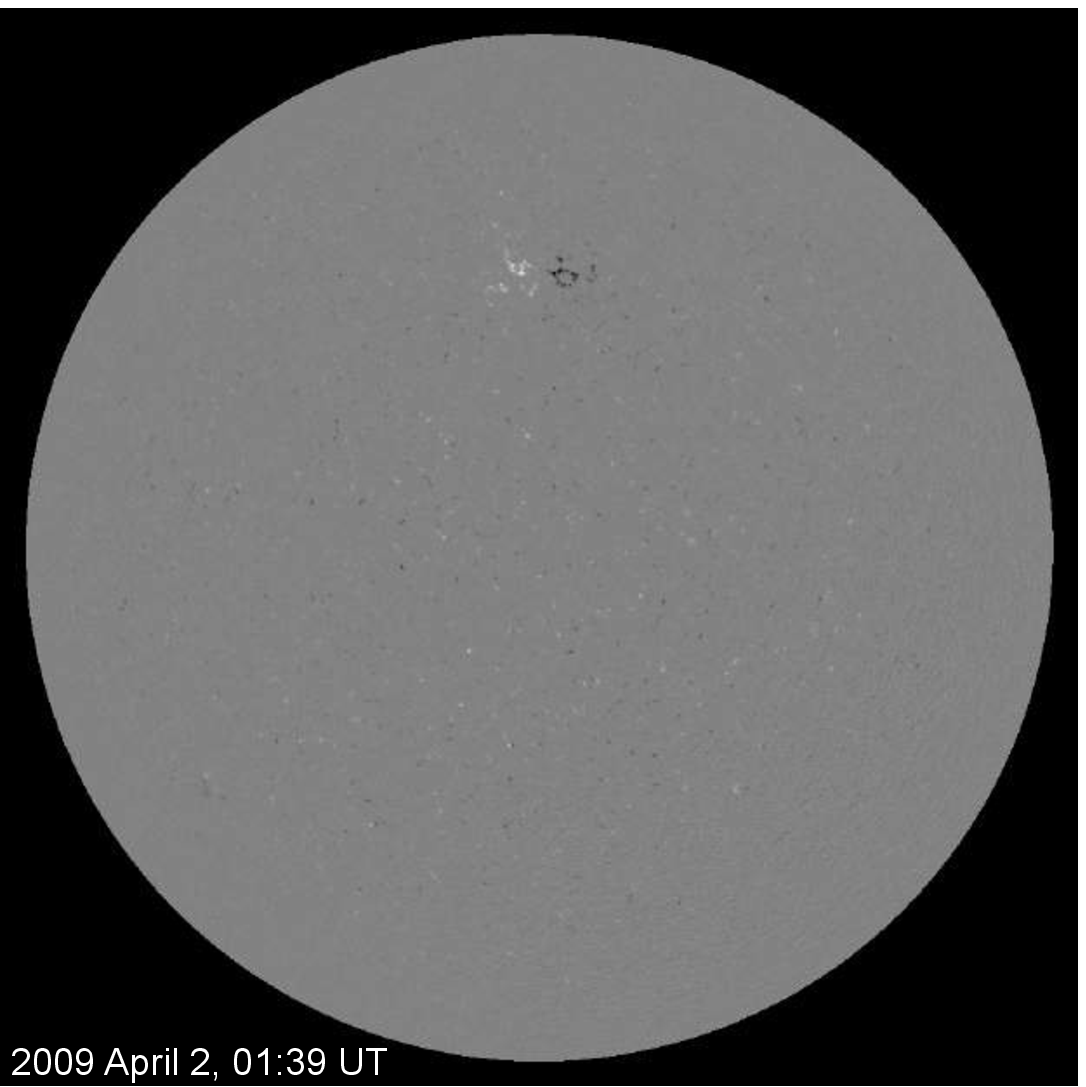}
\caption{Magnetogram taken by MDI on April 2, 2009.}
\label{F:MDI}
\end{figure}

In order to prove the correspondence between the field configuration in the breakout model (Figure~\ref{F:Breakout}) and the actual magnetic topology, we reconstructed the field in a potential approximation using MDI data. The magnetogram of the AR obtained on 2009 April 2, is presented in Figure~\ref{F:MDI}. As we can see, the AR is bipolar and no other strong AR was observed. The only unaccounted for source of the observed `quadrupolarity' is the global magnetic field of the Sun. We suggest that the two missing poles represent the north and south poles of the Sun. This assumption is justified because the studied AR was the only existing large-scale bipole on the Sun. 

To check this assumption, we applied the Potential Field Source Surface  model \citep[PFSS,][]{Mackay2012}, which is based on the spherical harmonic decomposition of the field. The required input parameter to this model is the value of the radial component of the field on the photosphere. Because it is impossible to measure the field throughout the whole solar surface at one time, we used a field map `stitched' from multiple MDI magnetograms scanned during the current solar rotation. PFSS has a single free parameter---the height of the source surface, i.e., the distance at which field potential is set to zero. For our modeling, we took the standard value of 2.5 solar radii. 

\begin{figure*}[th]
\centering
\includegraphics[width = \textwidth]{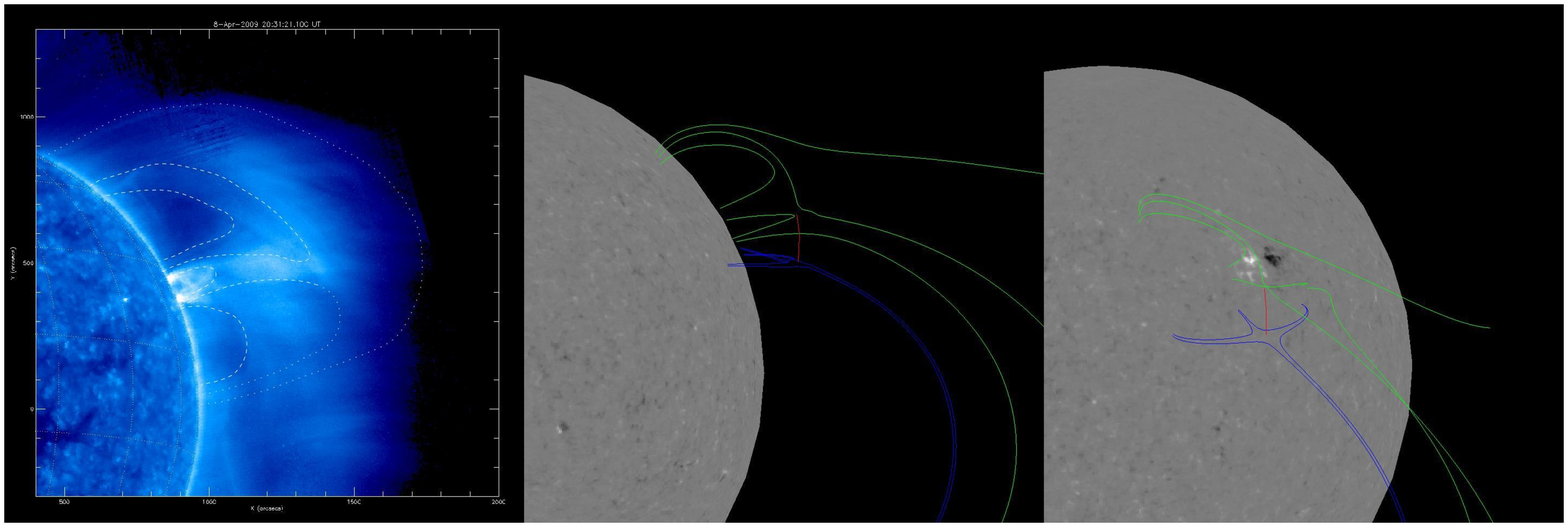}
\caption{Left: TESIS image; middle: reconstructed magnetic field, TESIS point of view; right: reconstructed magnetic field, behind-the-limb point of view. Blue: the field lines adjacent to the null-point; red: the separator field line originating from the null; green: other field lines.}
\label{F:Magnetic_field}
\end{figure*}

The reconstructed magnetic field is shown in Figure~\ref{F:Magnetic_field}. The notable feature of this configuration is a coronal null-point created by the positive source of AR and the south pole, which is situated at a height of approximately 120~Mm above the surface. The field lines adjacent to the null-point are blue in the images. The separator field line (shown in red) originating from the null divides the overall flux into four major domains connecting four given sources. Figure~\ref{F:Magnetic_field} shows only part of the separator, which connects the null-point and the highest point of the arcade.

The 3D field highly resembles the breakout model. The only significant difference is the X-point being substituted by the separator. Moreover, the side view of the 3D field almost perfectly fits the TESIS images (see Figure~\ref{F:Magnetic_field}), with the exception of a stretching factor of approximately two. If we scale the reconstructed magnetic field, the positions of the bright stripe and the separator coincide.

The discrepancy between the TESIS images and the reconstructed magnetic field can be explained by several reasons. First, since the pre-erupting structure had the energy to erupt, its magnetic field must be non-potential and therefore cannot coincide with the potential field obtained with the PFSS method. The second reason is the time lag between the magnetogram and the TESIS images. In fact, the part of the PFSS field map containing the magnetogram of the AR was taken at the moment of crossing the central meridian (see Figure~\ref{F:MDI}), i.e., in an early phase of its evolution, after which the AR strength may have increased. Finally, the deviation of the model parameters (such as the height of the source surface) and the contribution of high-order harmonics in the PFSS method could also affect the result.

\subsection{STEREO Observations}

\begin{figure*}[thb]
\centering
\includegraphics[width = 0.95\textwidth]{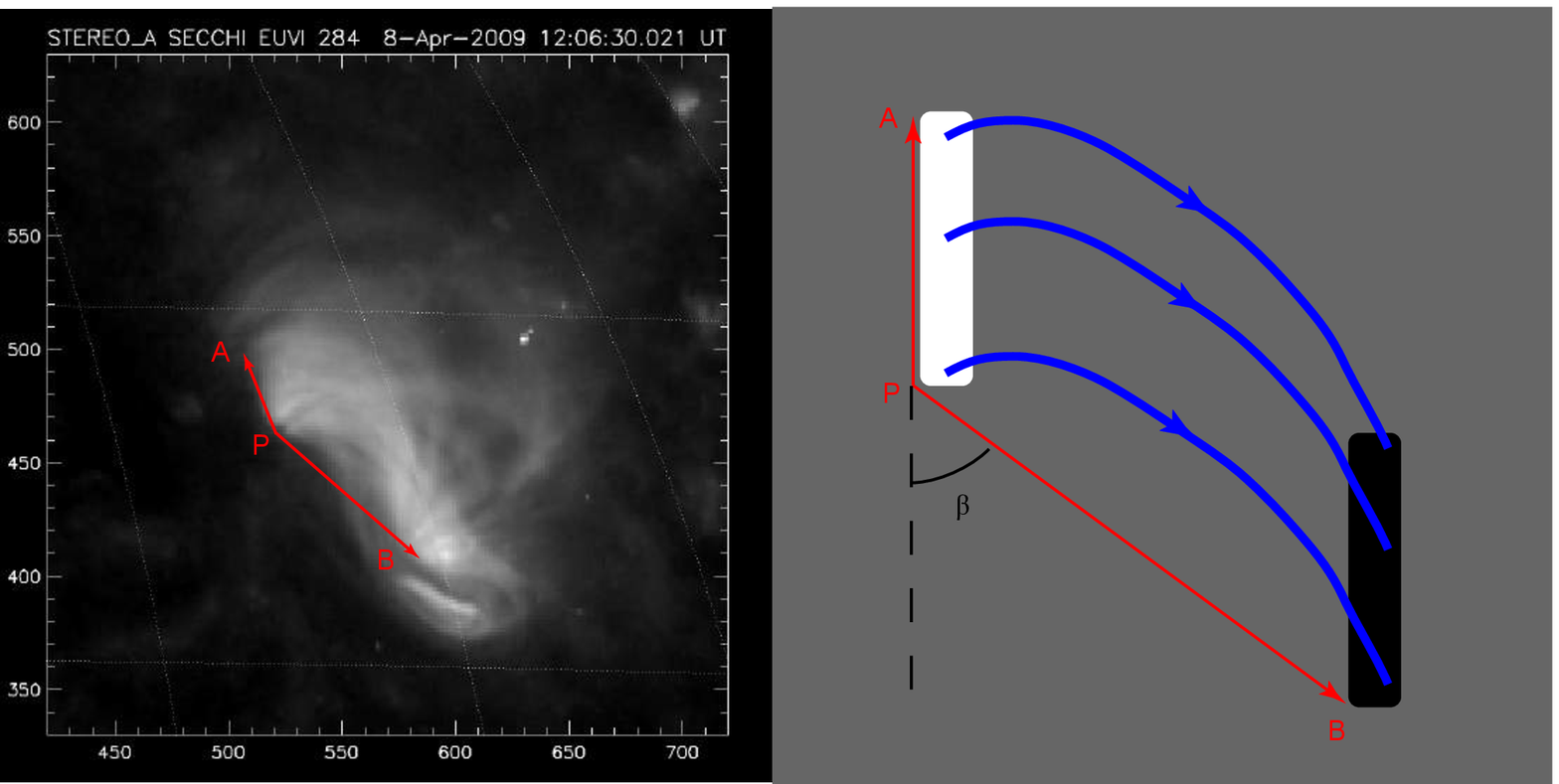}
\caption{Measurement of the shear angle. Left:EUVI-A 284~\AA\ image of the arcade. Right: a schematic image of the arcade. White indicates positive polarity; black---negative; gray---neutral.}
\label{F:shear_angle}
\end{figure*}

In the low corona, the CME pre-eruption structure was a sheared arcade, which was seen on the EUVI-A 284~\AA\ images (see Figure~\ref{F:shear_angle}). In other EUVI channels the sheared arcade was less distinctive. To determine a shear angle, we measured coordinates of the arcade footpoints  (points $P$, $A$, and $B$ in Figure~\ref{F:shear_angle}.) Assuming that these points lie on the solar surface, we obtained 3D coordinates of the vectors $\overrightarrow{PA}$ and $\overrightarrow{PB}$, and then determine the angle $\beta$ (see Figure~\ref{F:shear_angle}). To estimate error bars, we repeated this procedure 25 times. We obtained that $\beta = 40 \pm 4^{\circ}$, that is, the arcade is indeed sheared. 

\begin{figure*}[thb]
\centering
\includegraphics[width = 0.95\textwidth]{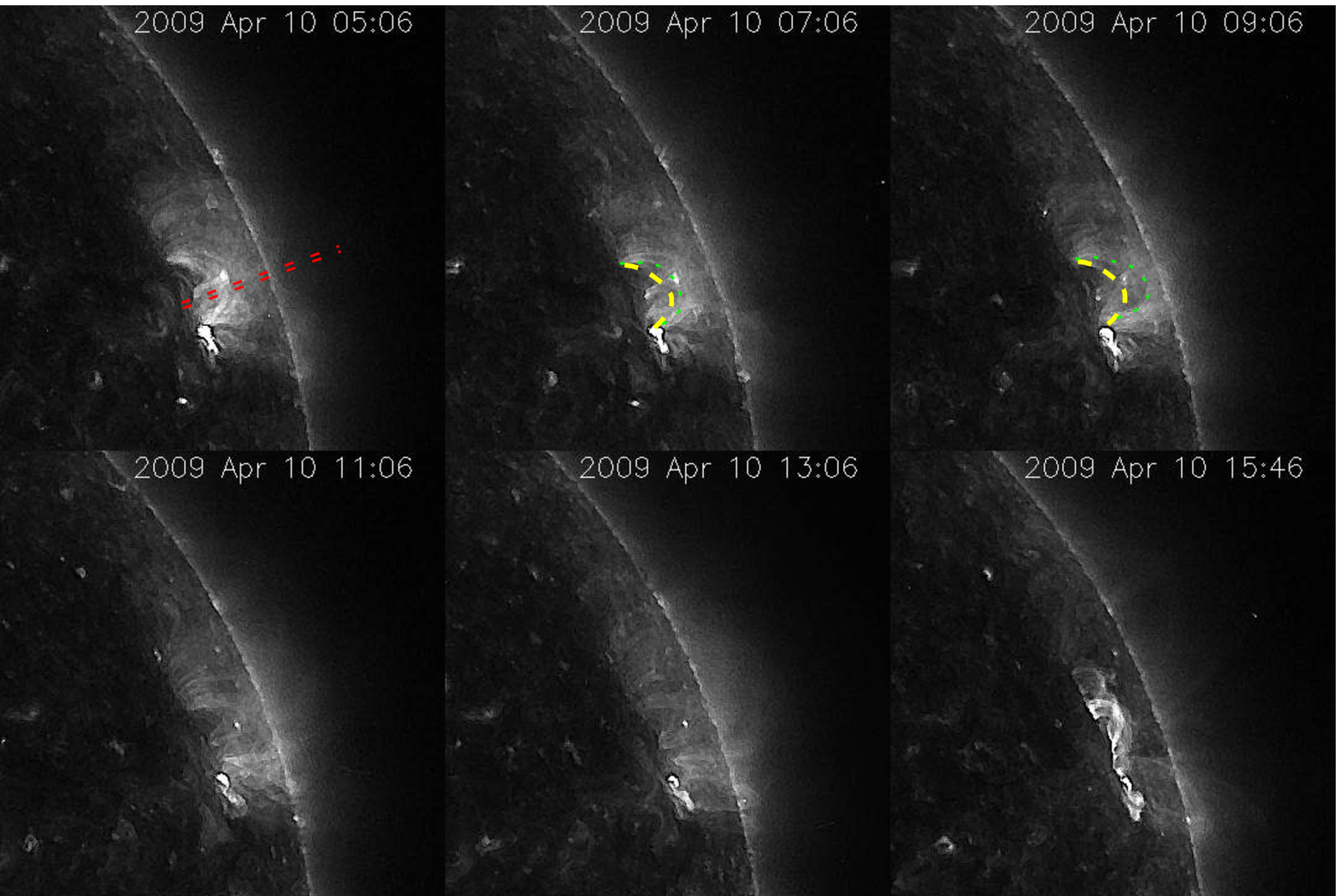}
\caption{Slow upward motion of the loops of the CME  arcade. These are sharpened \textit{STEREO-A} 284~\AA\ images, which were taken on April 10, 2009. Dashed yellow lines indicate positions of the loop at 05:06~UT. Green dotted lines indicate positions of the loop at the time the image was taken. Red dashed lines indicate the position of the slices, which were used in the construction of the Figure~\ref{F:stereo_slices}.}
\label{F:stereo284}
\end{figure*}

\begin{figure*}[thb]
\centering
\includegraphics[width = 0.95\textwidth]{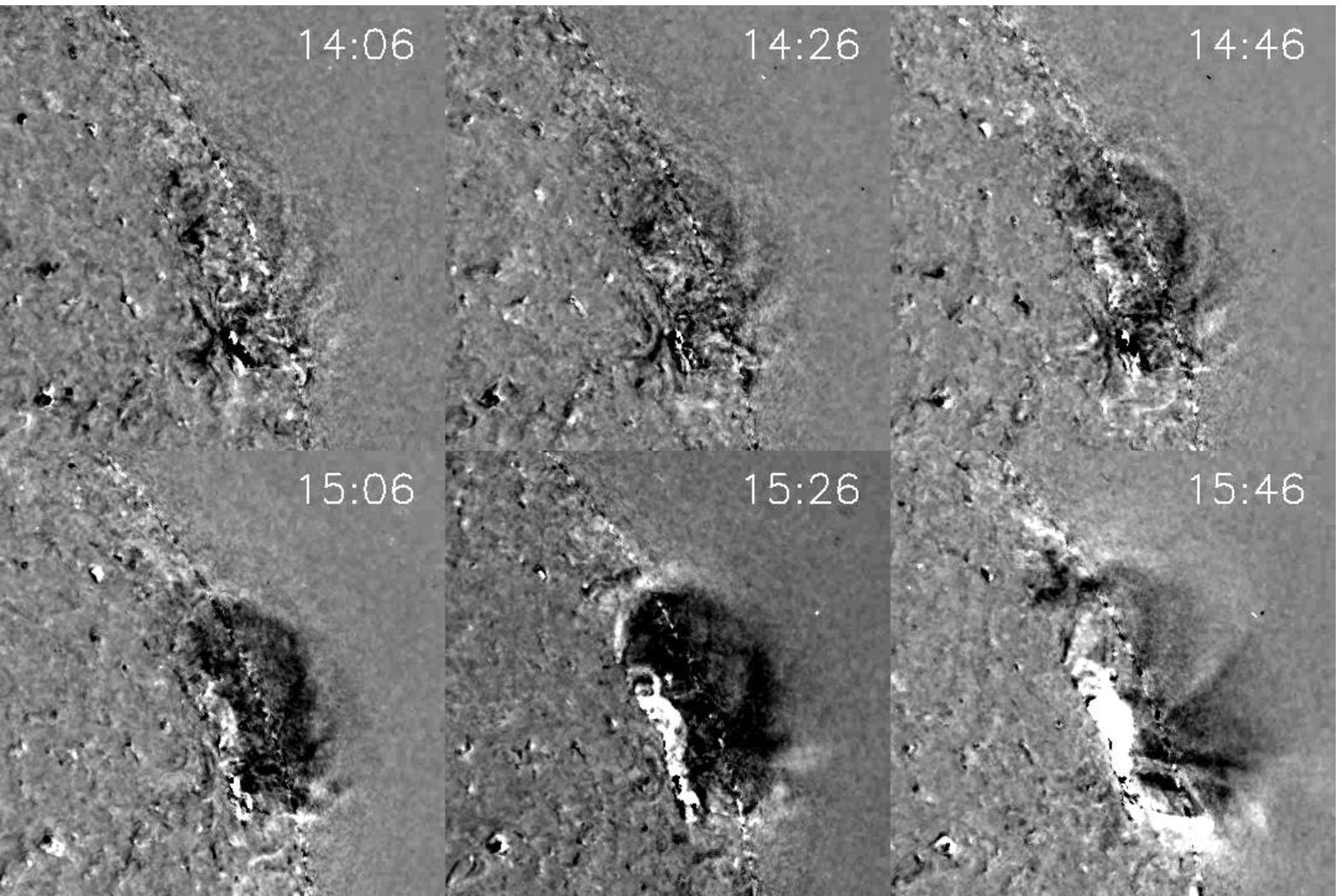}
\caption{\textit{STEREO-A} 284~\AA\ images running difference images. Images were taken on April 10, 2009.}
\label{F:stereo284_diff}
\end{figure*}

Loops of the arcade slowly moved up (see Figure~\ref{F:stereo284}). As the loops moved up, their intensity decreased. At 13:36~UT, the CME appeared on the EUVI running difference images (see Figure~\ref{F:stereo284_diff}). At 15:26~UT, a flare arcade appeared on the EUVI 284~\AA\ images.

To determine when the loops of the arcade started to move up, we made the stack-plot of slices, cut from running difference EUVI-A 284~\AA\ images. The positions of the slices are indicated in Figure~\ref{F:stereo284}. All slices were cut in the radial direction. The position of the slices were differentially rotated for each image, in order to exclude rotation of the Sun from the analysis.

\begin{figure*}[thb]
\centering
\includegraphics[width = 0.95\textwidth]{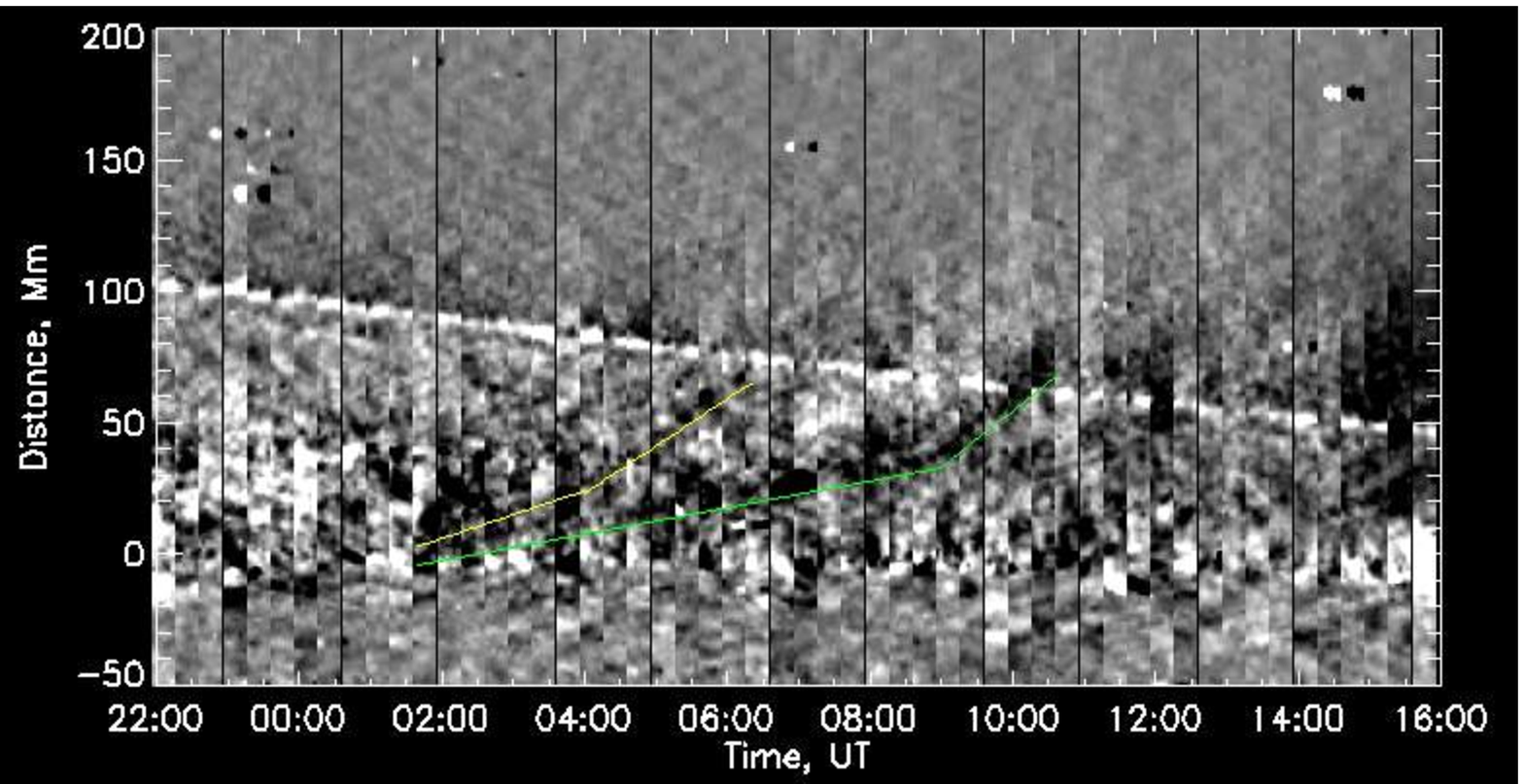}
\caption{Stack-plot of slices, which were cut from running difference images of STEREO-A 284~\AA. Green and yellow lines indicate loops moving upwards. Position of the slices is indicated in Figure~\ref{F:stereo284}. The width of the slices is 10$^{\prime\prime}$ (7.2~Mm).}
\label{F:stereo_slices}
\end{figure*}

On the obtained stack-plot, we see several dark linear features (see Figure~\ref{F:stereo_slices}). These features correspond to the expanding loops. We marked the two most distinctive features with green and yellow lines.  Their speeds lie in the range of 1--6~km~s$^{-1}$ (values are obtained from linear fitting). The motion started at $\approx$~01:45~UT (almost at the same time that the bright stripe appeared).

\begin{figure}[thb]
\centering
\includegraphics[width = 0.45\textwidth]{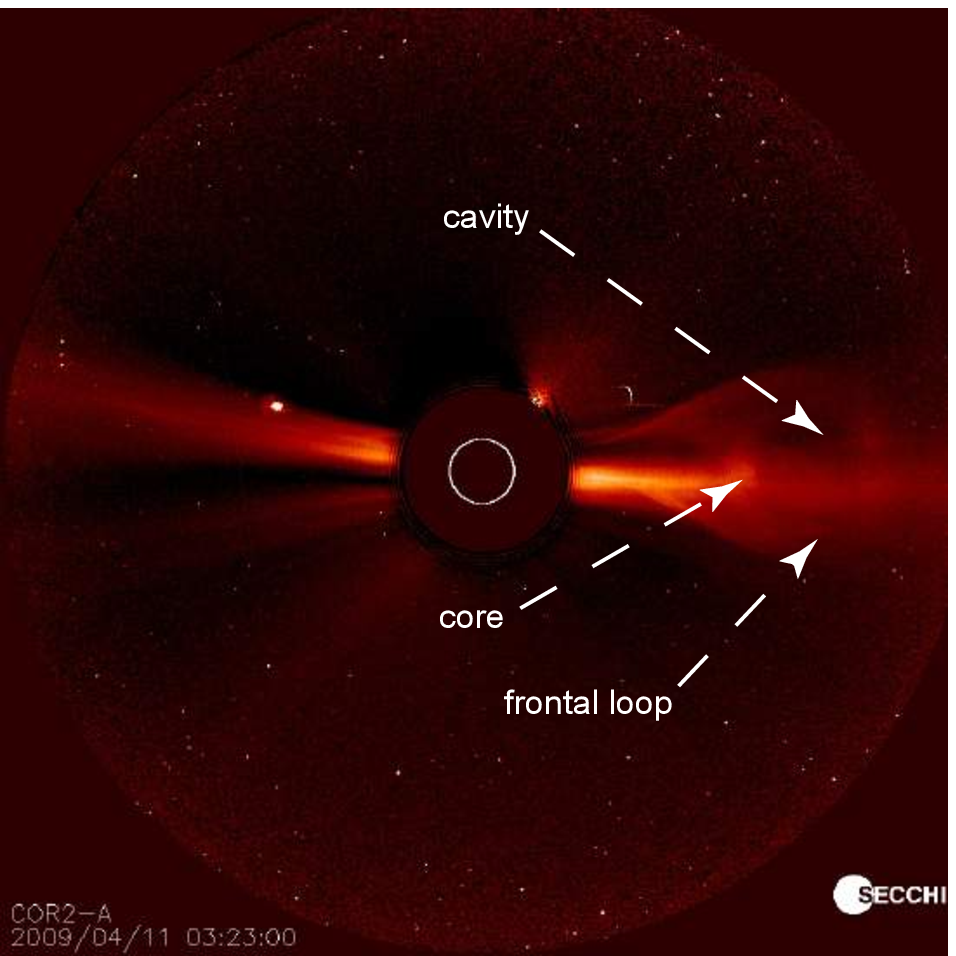}
\caption{CME three-part structure.}
\label{F:3part_structure}
\end{figure}

On the COR2 images, the CME had a three-part structure: bright core, dark cavity, and bright frontal loop (see Figure~\ref{F:3part_structure}). The core, which is usually interpreted as a prominence, was not seen on the TESIS  and EUVI images.

\subsection{Kinematics}

The CME was observed by TESIS, EUVI, COR2, and LASCO. These instruments cover different distance ranges, and it is possible to study CME kinematics from the solar surface up to the boundaries of the LASCO C3 field of view. However, this research should be carried out with caution. TESIS and \textit{STEREO-A} EUVI observe thermal emission of plasma with different temperatures, while COR2 and LASCO observe scattered light. We should be careful to ensure that we measure kinematics of the same part of the CME. In this research, we will study kinematics of the CME frontal loop.

\begin{figure}[!thb]
\centering
\includegraphics[width = 0.45\textwidth]{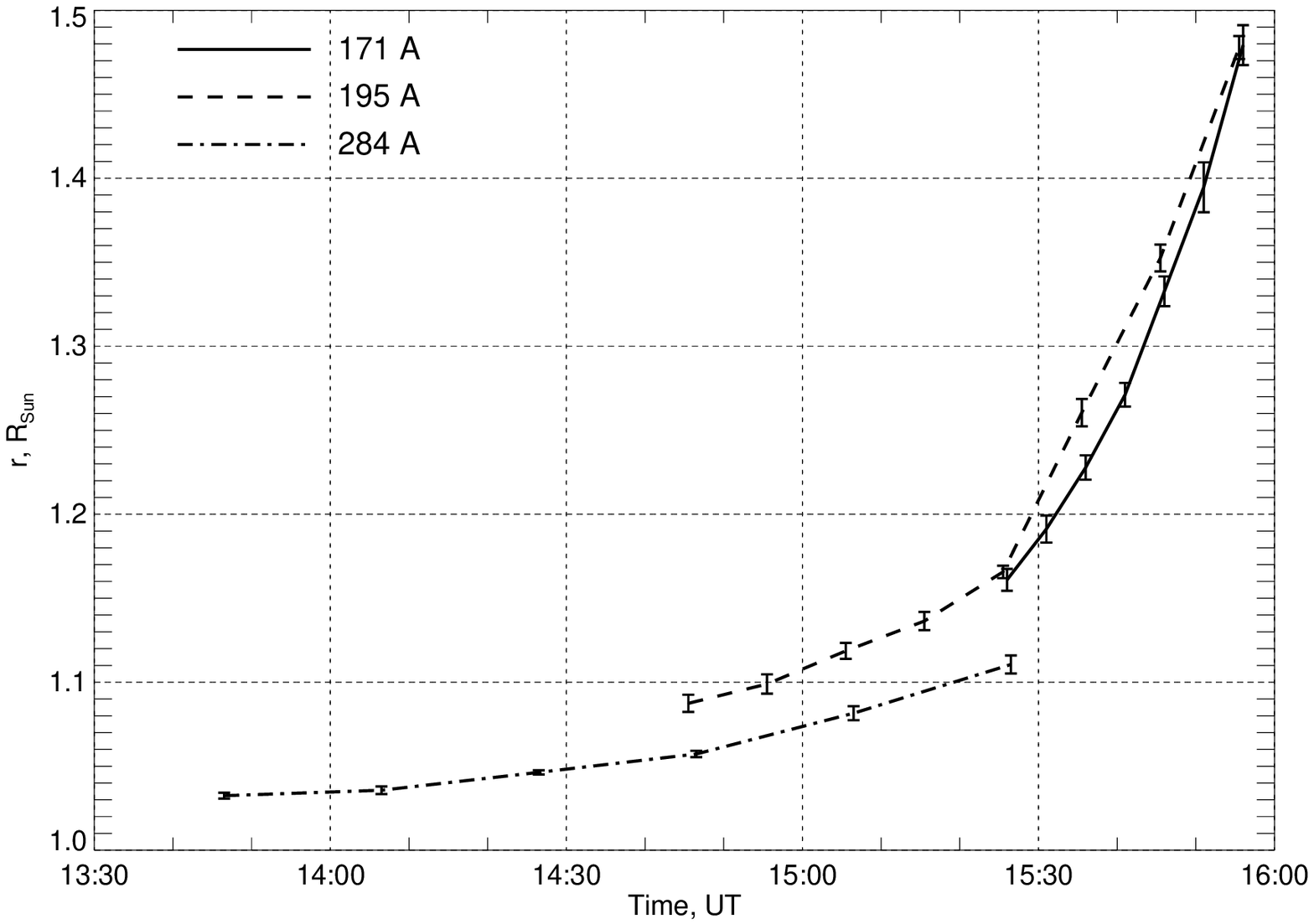}
\caption{Radial coordinates of the front edges in different \textit{STEREO-A} EUVI channels. Solid line: 171~\AA\ channel; dashed line: 195~\AA\ channel; dashed dotted line: 284~\AA\ channel.}
\label{F:stereo_kinematics_raw}
\end{figure}

We used running difference images to measure frontal loop coordinates on the TESIS and EUVI images, and normal images on LASCO and COR2. The measurement was carried out by a simple point-and-click procedure. To estimate error bars, we repeated the procedure 9~times. The result of the kinematics measured on EUVI images is shown in Figure~\ref{F:stereo_kinematics_raw}. The curves in different channels deviate from one another because different channels have different temperature response functions and we measure slightly different parts of the CME.

\begin{figure}[thb]
\centering
\includegraphics[width = 0.45\textwidth]{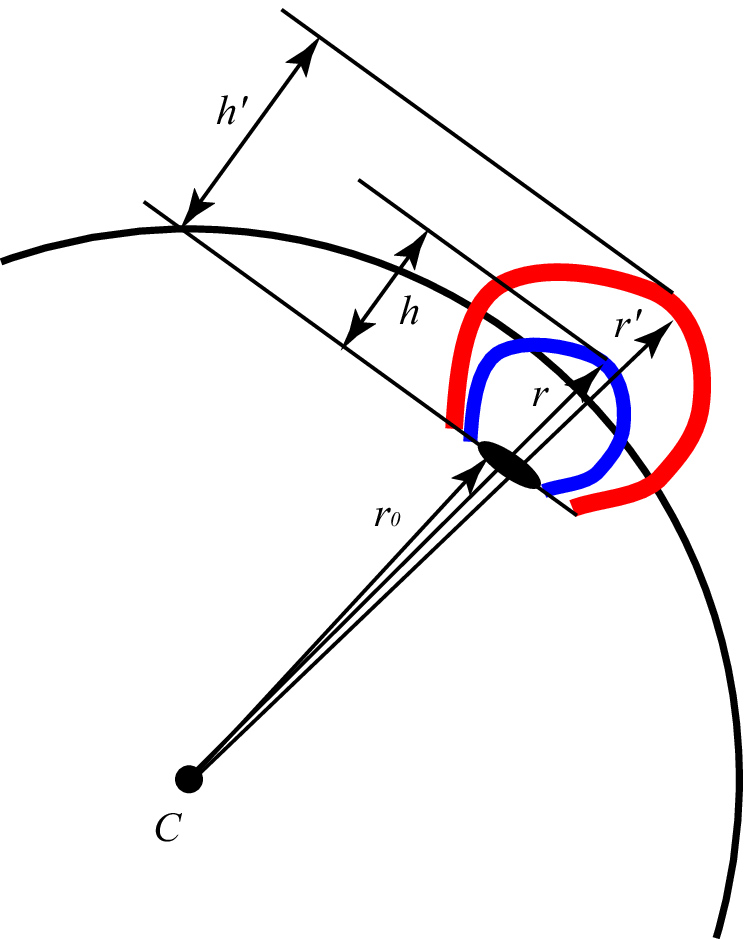}
\caption{Scaling of different EUVI channels.}
\label{F:EUVI_scaling}
\end{figure}

To fix this deviation, we propose the following model. We assume that all parts of the CME expanded proportionally. If $h(t)$ and $h^{\prime}(t)$ are heights of different parts of the CME (see Figure~\ref{F:EUVI_scaling}), we assume that
\begin{equation}
h^{\prime}(t) = \alpha h(t),
\label{E:h}
\end{equation}
where $\alpha = const$. We can rewrite equation (\ref{E:h}) using radial coordinates (distance from the Sun's center on the image):
\begin{equation}
r^{\prime} - r_0 = \alpha (r - r_0)
\end{equation}
\begin{equation}
r^{\prime}  = \alpha (r - r_0) + r_0,
\label{E:r_prime}
\end{equation}
where $r_0$ is the radial coordinate of the CME footpoints, and $r^{\prime}$ and $r$ are radial coordinates of CME parts observed in different channels. We scaled channels 284~\AA\ and 171~\AA\ using formula~(\ref{E:r_prime}) to match the curve of the 195~\AA\ channel. For 284~\AA\ we used $\alpha = 1.26$, and $\alpha = 1.05$ for 171~\AA. The scaled curve fit the 195~\AA\ curve well, which shows that our model is reasonable.

\begin{figure*}[!thb]
\centering
\includegraphics[width = 0.91\textwidth]{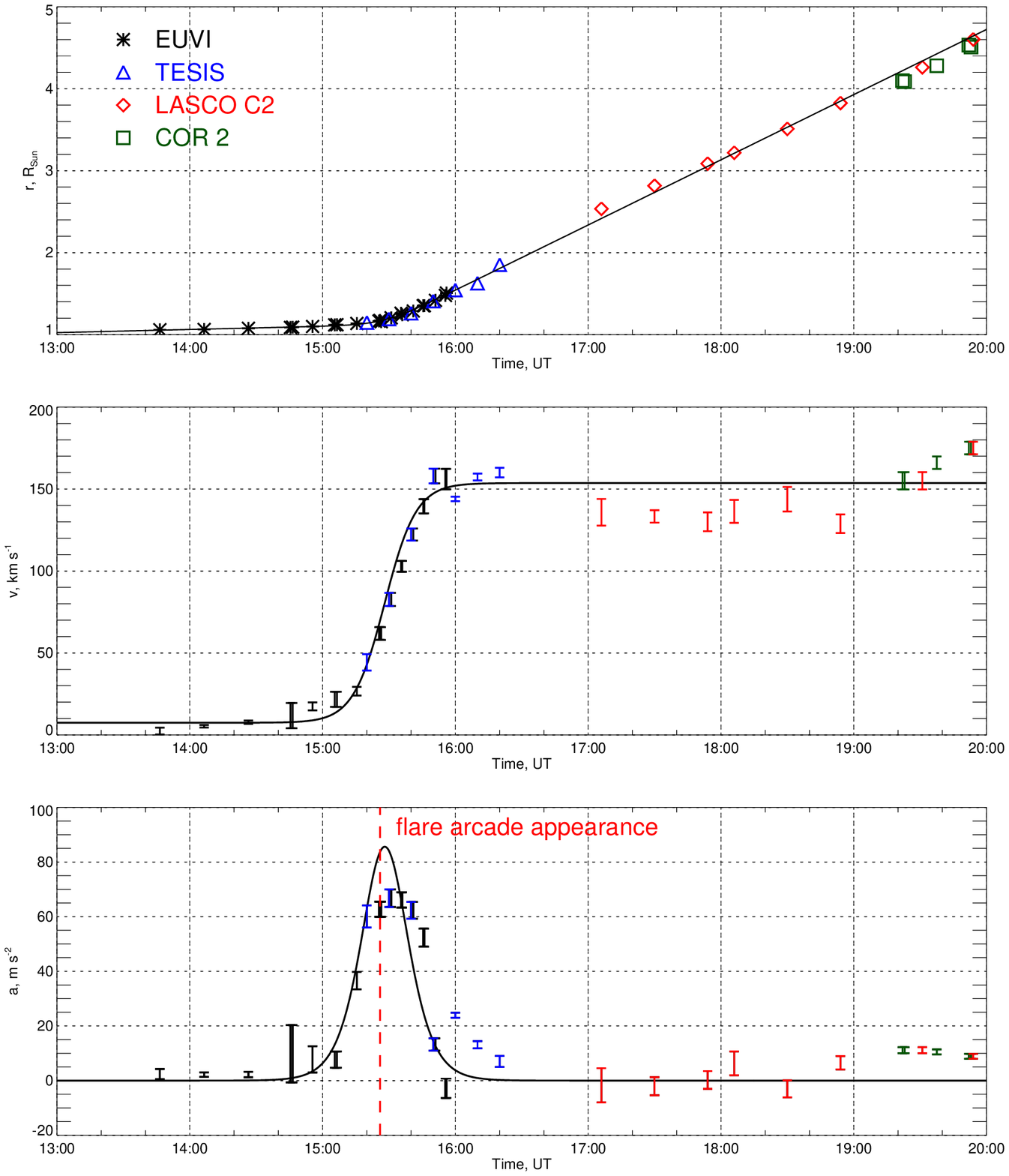}
\caption{Combined CME kinematics. Top: radial coordinate of the CME front edge; middle: radial velocity; bottom: radial acceleration. Black stars: \textit{STEREO-A} EUVI channels; blue triangle: TESIS EUV telescope; red diamond: LASCO C2; green square: \textit{STEREO-A} COR2. Vertical dashed red line indicates appearance of flare arcade on the \textit{STEREO-A} 284~\AA\ images.}
\label{F:kinematics_small}
\end{figure*}

\begin{figure*}[!thb]
\centering
\includegraphics[width = 0.91\textwidth]{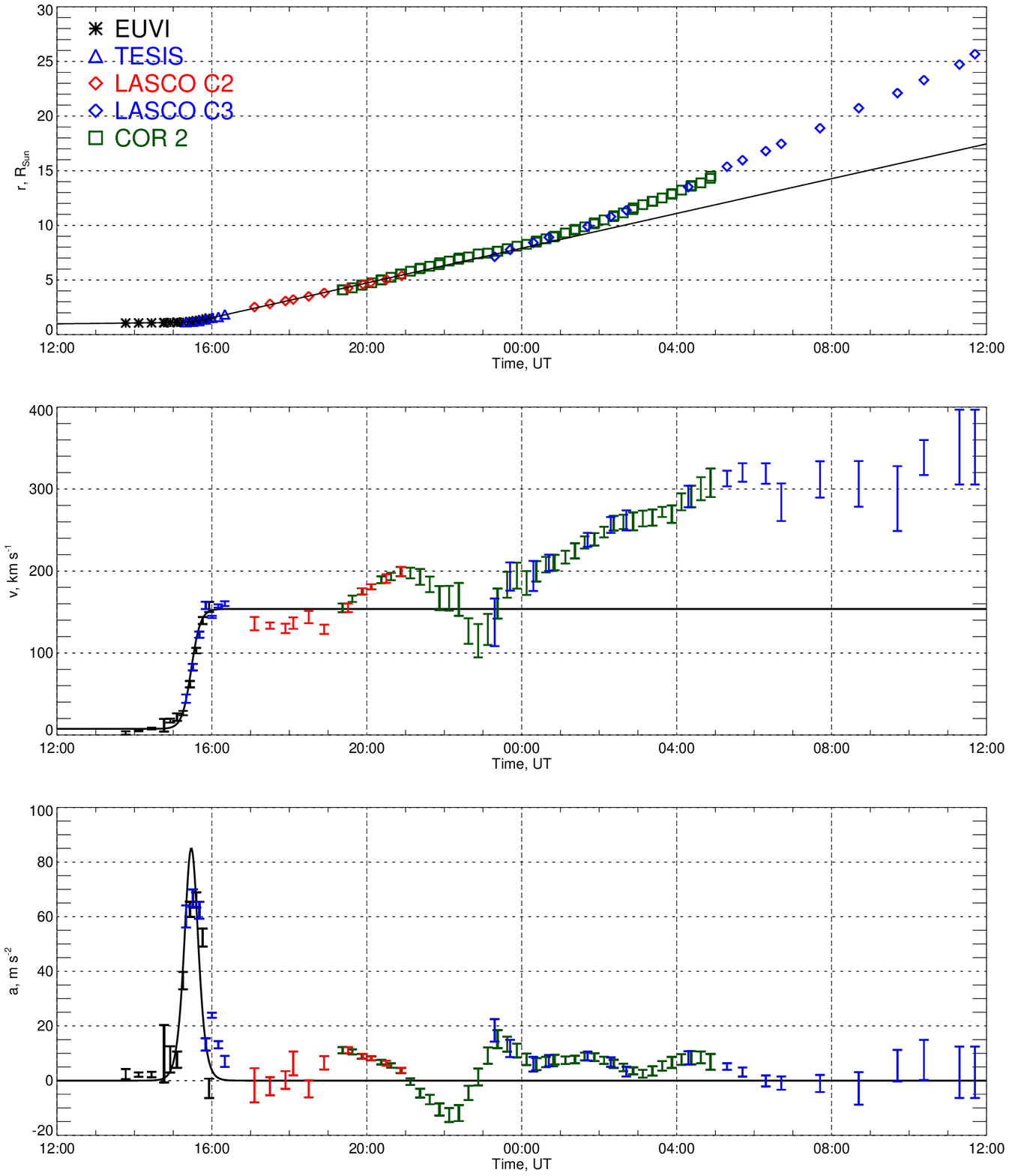}
\caption{Combined CME kinematics. Top: radial coordinate of the CME front edge; middle: radial velocity; bottom: radial acceleration. Black stars: \textit{STEREO-A} EUVI channels; blue triangle: TESIS EUV telescope; red diamond: LASCO C2; blue diamond: LASCO C3; green square: \textit{STEREO-A} COR2.}
\label{F:kinematics_big}
\end{figure*}

The TESIS and LASCO line of sight was inclined by 46$^{\circ}$ to the \textit{STEREO-A} line of sight (see Figure~\ref{F:stereo_position}). The CME trajectory was inclined by approximately the same angle  to the TESIS/LASCO and \textit{STEREO-A} image planes. That is why TESIS and LASCO kinematics curves almost coincide with EUVI and COR2 curves. To make the correspondence better, we multiplied TESIS coordinates by 0.99, LASCO/C2 by 1.02, and LASCO/C3 by 1.11. This correction accounts for different projection angles and the fact that different instruments observed different parts of the CME. After the correction, all data fit perfectly. We numerically differentiated the height-time profile and obtained CME velocity and acceleration (see Figures \ref{F:kinematics_small} and \ref{F:kinematics_big}).

To determine timing of the CME acceleration, we approximated $r(t)$ with the formula \citep{Sheeley2007}
\begin{equation}
r(t) =  r_1 + \frac{v_f + v_0}{2}(t - t_1) + \frac{v_f-v_0}{2}\tau \ln \cosh \left(\frac{t-t_1}{\tau}\right),
\label{E:r_t}
\end{equation}
where $t_1$---time of the maximum acceleration; $\tau$---acceleration timescale; $v_0$---initial CME velocity; $v_f$---final CME velocity; and $r_1$---CME position at $t = t_1$ ($r_1 = r(t_1)$).

We differentiated equation~(\ref{E:r_t}) and calculated velocity and acceleration. The result of the approximation is plotted in Figures \ref{F:kinematics_small} and \ref{F:kinematics_big}. As we see, formula~(\ref{E:r_t}) describes kinematics well up to 5~$R_\odot$. The CME is impulsively accelerated from 15:10 until 15:50~UT. The peak acceleration was 85~m~s$^{-2}$, the velocity before impulsive acceleration was 7~km~s$^{-1}$, and the velocity after impulsive acceleration was 155~km~s$^{-1}$. During impulsive acceleration, a flare arcade appeared on the EUVI 284~\AA\ images (see Figures \ref{F:kinematics_small} and \ref{F:stereo284_diff}).

At distances above 7~$R_\odot$ the CME accelerated up to 350~km s$^{-1}$ over $\approx$~7~hours with acceleration $\approx$~10~m~s$^{-2}$. 

The CME trajectory was inclined by approximately 23$^{\circ}$ to image planes. This means that real values of velocities and accelerations should be 8~\% higher.

\subsection{Summary of the Observations}

We will briefly summarize the observations.
\begin{enumerate}
\item EUVI-A 284~\AA\ observed a sheared arcade.
\item From 12:00~UT April~8 until 21:45~UT on 2009 April~9, TESIS observed quadrupolar structure.
\item A reconstructed from the MDI data magnetic field has a multipolar structure with an X-point above the AR.
\item From 21:45~UT April~9 until 01:15~UT on 2009 April~10, loops near the X-point moved away from each other.
\item At 01:15~UT, a bright stripe appeared between the loops. At 04:15~UT, the bright stripe disappeared.
\item From 01:45 to 04:30~UT, the flux in the GOES 0.5--4~\AA\ channel increased.
\item At 01:45~UT, loops below the X-point started to slowly move up on the EUVI-A 284~\AA\ images.
\item From 15:10~UT to 15:50~UT, the CME impulsively accelerated.
\item At 15:26~UT, a flare arcade appeared on the EUVI-A 284~\AA\ images.
\item From 15:50~UT to 19:00~UT, the CME moved with constant velocity.
\item After 19:00~UT, the CME slowly accelerated up to 350~km~s$^{-1}$.
\end{enumerate}

\begin{figure*}[!thb]
\centering
\includegraphics[width = 0.85\textwidth]{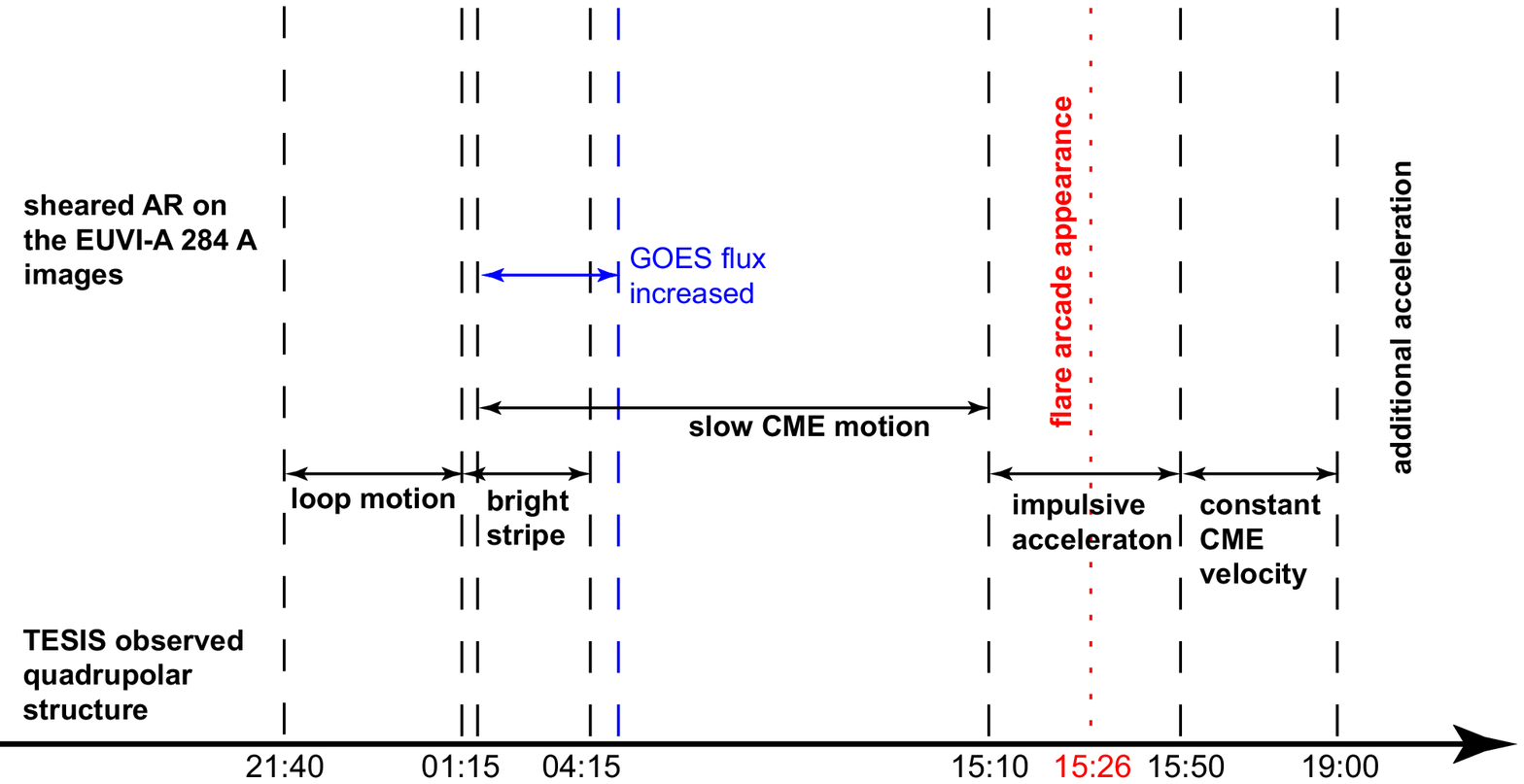}
\caption{Timing of the observations.}
\label{F:Timing}
\end{figure*}

The timing of the observations is illustrated in Figure~\ref{F:Timing}.

\section{Discussion}

\subsection{Breakout Model and Observations}

We presented observations of the CME in the far corona, which showed evidence of the breakout model. Below we will compare the breakout scenario with the observations.

\begin{enumerate}
\item    \emph{Model}: The CME pre-eruption structure should be quadrupolar. 
\newline \emph{Observations:} The AR  pre-eruptive configuration was quadrupolar on the TESIS Fe~171~\AA\ images. Since  magnetic field lines are lightened on the Fe~171~\AA\ images, we interpret the observed structure as a magnetic quadrupolar structure. Also, a reconstructed from the MDI data magnetic field has a multipolar structure with an X-point above the AR.
\item    \emph{Model}: The driving mechanism for the eruption is the shearing motion of the arcade footpoints, which lie below the X-point. 
\newline \emph{Observations:} We observed a sheared arcade on the EUVI-A 284~\AA\ images. The sheared arcade is seen most distinctively during and after the CME impulsive acceleration.
\item    \emph{Model}: The shearing arcade motion should eventually lead to the stretching of the X-point into the current sheet. When the current sheet is stretched enough, the breakout reconnection occurs.
\newline \emph{Observations:} We see that loops, which formed the X-point, moved away from each other. The movement of the loops near the X-point stopped for $\approx$~3~hours, a bright stripe appeared between them, and a GOES flux in the 0.5--4~\AA\ channel increased. We interpret the loop motion and the stripe as evidence of the breakout reconnection.
\item    \emph{Model}: After the breakout reconnection, the CME  should  start to slowly move up. 
\newline \emph{Observations:} We see slow upward loop motion on the EUVI-A 284~\AA\ images.
\item    \emph{Model}: The slow CME  motion should stretch a current sheet below it. This will cause a flare and an impulsive CME acceleration. 
\newline \emph{Observations:} We see simultaneous impulsive CME  acceleration and a flare arcade around 15:26~UT on the EUVI-A 284~\AA\ images.
\item    \emph{Model}: After the impulsive acceleration ends, CME  should move with constant velocity. 
\newline \emph{Observations:} After the impulsive acceleration, the CME moved with constant velocity.
\end{enumerate}

The comparison is summarized in Table~\ref{T:model_observation}. The observed CME  explicitly followed the breakout model scenario: quadrupolar structure, sheared arcade, breakout reconnection, simultaneous impulsive acceleration and a flare, and the CME kinematics. The main result of our work is the observation of the key element of the model---the breakout reconnection.

\begin{table*}[bht]
\centering
\caption{Comparison of the breakout model and the observations}
\begin{tabular}{lll}
\hline
Model                                       & Observations                                      & Instrument       \\
\hline
Quadrupolar pre-erupting magnetic structure & Quadrupolar pre-erupting structure                & TESIS Fe~171~\AA \\
                                            & Reconstructed magnetic field                      & MDI              \\
Shearing arcade motion                      & Sheared arcade                                    & EUVI-A 284~\AA   \\
Breakout reconnection                       & Loops moved away from each other near the X-point & TESIS Fe~171~\AA \\
                                            & Bright stripe                                     & TESIS Fe~171~\AA \\
Slow CME acceleration                       & Slow loop upward motion                           & EUVI-A 284~\AA   \\
Flare                                       & Post-flare arcade                                 & EUVI-A 284~\AA   \\
Impulsive CME acceleration                  & Impulsive CME acceleration                        & Kinematics       \\
Propagation with a constant velocity        & Propagation with a constant velocity              & Kinematics       \\
\hline
\end{tabular}
\label{T:model_observation}
\end{table*}

\subsection{Bright Stripe}

In flares during reconnection, plasma is heated up to 10~MK. Such hot plasma should be seen as a darkening in the 171~\AA\ line. In our observations, however, we see brightening. We found two ways to explain why the stripe is bright, but not dark.

The loops near the X-point moved slowly, and corresponding reconnection should also be slow. The magnetic field of the AR was weak. In these circumstances, the reconnection rate could be sufficient to heat plasma to temperatures of only around 1~MK. In this case, the current sheet should be seen as a brightening in the 171~\AA\ line.

An alternative explanation is that we see an increase in density. During reconnection, plasma inflows from areas surrounding the current sheet increase the current-sheet density. Also, at high altitudes, scattering---which depends on density---contributes to the emission in the 171~\AA\ line \citep{Reva2014}. It is possible that the observed bright stripe is the scattered light in the dense current sheet.

There are also few observations in which a brightening in the 171~\AA\ line was interpreted as a reconnection. \citet{Masson2014} analyzed the pseudo-streamer observed by Atmospheric Imaging Assembly \citep[AIA][]{Lemen2011}, and \citet{Driel2014} analyzed an active region near the area of the expanding CME observed by AIA. Both authors reconstructed magnetic fields using Helioseismic and Magnetic Imager data (HMI, \citet{Schou2012}), and showed that there was a brightening in the 171~\AA\ line near the magnetic null-point. Both authors interpreted brightening as a reconnection in the current sheet.

Although the bright stripe looks like a current sheet, it could occur due to other reasons: plasma compression; plasma heating or cooling to the formation temperature of the Fe~171~\AA\ line; or a chance alignment of several structures along the line of sight during the magnetic field reconfiguration. However, the bright stripe was accompanied by other evidence of the reconnection---the sideways loop motion. The bright stripe could be explained in the reconnection framework. There are observations where brightening in the 171~\AA\ line was interpreted as a reconnection in the current sheet. We think that our interpretation of the bright stripe as a reconnection in the current sheet is a reasonable conclusion.

\subsection{GOES Signal}

When the bright stripe appeared, the GOES flux increased. Although, it is tempting to  interpret the GOES signal as an indirect evidence of the breakout reconnection, we cannot prove the connection between the bright stripe and the increase in the GOES flux. GOES do not have imaging capabilities, and the flux increase could be just a coincidence. The signal was very weak, and it could come from other parts of the Sun.  However, even if the GOES signal was not related to the breakout reconnection, it is still a useful information. It provides a higher estimate on the possible soft-X-ray flux from the breakout reconnection region.

During eruption, the source AR for the analyzed CME was behind the solar limb, as seen from the terrestrial viewpoint. This makes it impossible to use the GOES data to find the class of the associated flare.

\subsection{Timing of the CME and the Breakout Reconnection}

According to our observations, the breakout reconnection occurred from 01:15~UT to 04:15~UT, and the impulsive acceleration began at 15:10~UT. Such a long delay (11~hr) looks unusual and deserves discussion. In this section, we will compare the delay with previous observations of the evidence of the breakout reconnection, numerical simulations, and kinematics timing.

\citet{Manoharan2003, Aurass2011, Aurass2013} observed evidence of breakout reconnection in radio data 1--5~minutes before the CME. \citet{Gary2004} observed brightenings (which authors interpreted as a breakout reconnection) 4~minutes before the CME. \citet{Sterling2001} observed `EIT crinkles' 40--50~minutes before the CME. \citet{Sterling2004Feb} observed pre-eruption dimming 6~hr before CME. We see that, in previous observations, the delay between the breakout reconnection and the CME ranges from a few minutes to 6 hr.

In numerical simulations, the delay between breakout reconnection and impulsive CME acceleration encompasses a wide range of possible values: 10~minutes in \citet{Lynch2005}, 15~minutes in \citet{Masson2013}, 8~hr in \citet{Karpen2012}, and 13~hr in \citep{Zuccarello2008}. The delay observed in our work is consistent with the delay in the numerical simulations.

There is also another indirect way to check the reasonableness of the observed delay. In the breakout model, the CME has three phase kinematics: slow acceleration (after breakout reconnection and before flare reconnection), impulsive acceleration (during flare reconnection), and propagation with constant velocity (after flare reconnection). If we assume that the observed in CMEs three phase kinematics is caused by the breakout scenario, then we can use the delay between the CME onset and the start of the impulsive acceleration as an estimate of the delay between breakout and flare reconnections. \citet{Alissandrakis2013} reported a 15-minutes delay, \citet{Zhang2001} reported a 1--1.5-hr delay, \citet{Lin2010} reported a 4-hr delay, and \citet{Reva2014} reported a 12-hr delay. The delay in kinematics phases ranges from 15~minutes to 12~hours.

As we see, the delay between the breakout reconnection and the impulsive acceleration observed in our work is consistent with previous observations, numerical simulations, and CME kinematics timing. The delay ranges from several minutes to 13 hr. We think that the value of the delay is determined by the rate at which the energy is pumped into the system. The observed CME occurred during the deep minimum of the solar cycle. We think that the delay was long because at that time the energy pumping rate was low. Also, since for some CMEs breakout reconnection occurs few hours before eruption, the TESIS-like observations of the breakout reconnection could be used to forecast CMEs.

\subsection{Additional Acceleration}

Below 5~$R_\odot$ the CME had three phase kinematics, which is consistent with the breakout model. Above 5~$R_\odot$ the CME experienced additional acceleration from 150 to 350~km~s$^{-1}$. The breakout model describes the CME initiation, and the details of the CME propagation lie beyond the scope of the breakout model \citep{Antiochos1999, Lynch2008}. We need a different model to explain the additional CME acceleration.

\citet{Yashiro2004} showed that CMEs with a velocity greater than 400~km~s$^{-1}$ decelerate, and CMEs with a velocity less than 400~km~s$^{-1}$ accelerate. The authors proposed that fast CMEs move faster than the solar wind, and the wind decelerates them, while slow CMEs move slower than the solar wind, and the wind accelerates them.

The analyzed CME after breakout reconnection moved with a velocity 150 km~s$^{-1}$. It is less than solar wind velocity (400~km~s$^{-1}$), so the wind should accelerate the CME. Furthermore, the asymptotic velocity of the CME is 350~km~s$^{-1}$, which is close to the solar wind velocity.  We think that the additional acceleration of the analyzed CME was caused by the solar wind.

The breakout model describes CME initiation. The model of the CME acceleration by the solar wind describes CME propagation. Both models have their limitations: the breakout model cannot explain CME late evolution, while acceleration by the solar wind cannot explain CME initiation. Therefore, it is natural that the breakout model does not explain every aspect of the presented observations.

\subsection{Data We Did Not Use}

In this work, we used data from \textit{STEREO-A}, TESIS, MDI, LASCO, and GOES. We tried to adapt data from other instruments to our research, but did not include them in the paper. Below we will list these instruments and explain why we excluded them from the research.

We wanted to find another independent way to prove the presence of the breakout reconnection. We looked into radio data. During the breakout reconnection, the Sun was observed by the Hiraiso Radio Spectrograph \citep{Kondo1995}. The radio intensity slightly increased from 00:00~UT to 03:00~UT on April 10, which could be interpreted as a sign of the breakout reconnection. However, we find this conclusion too speculative, and that is why we did not include radio data in our research.

It would be great to have RHESSI \citep{Lin2002} imaging observations for the analyzed event. However, the X-ray flux from the Sun was very weak, and the RHESSI signal did not exceed the sensitivity threshold.

TESIS  far corona images are available only in the Fe~171~\AA\ line. It would be nice to complement TESIS images with far corona images in other wavelength. Sadly, other EUV and soft X-ray telescopes---for example, EIT \citep{del95} and XRT \citep{Golub2007}---do not have enough dynamic range to register far corona. TESIS Fe~171~\AA\ images were the only option to observe this event in the far corona.

\section{Conclusion}

We presented CME  observations with a TESIS Fe~171~\AA\ telescope. The telescope could observe corona up to 2~$R_\odot$. It observed an AR with a quadrupolar structure. In this structure, loops, which formed an X-point, started to move away from each other. Some time later, a bright stripe appeared between the loops. When the bright stripe disappeared, the loops below the  X-point started to slowly move up. Some time later, the CME erupted and a flare arcade formed below it.

We interpreted this event in terms of the breakout model. We think that quadrupolar pre-erupting structure is a breakout pre-CME structure, and loop motions near X-point and the bright stripe are evidence of the breakout reconnection. CME evolution precisely followed the breakout scenario. In our opinion, the presented observations are valuable evidence of the breakout model.

\acknowledgments
We are grateful to Brigitte Schmieder  for her valuable advice. This work was  supported by a grant from the Russian Foundation of
Basic Research (grant 14-02-00945) and by Program No. 9 for fundamental
research of the Presidium of the Russian Academy of Sciences.

\bibliographystyle{apj}
\bibliography{mybibl}

\end{document}